\documentclass[aps,twocolumn,prl,superscriptaddress,showpacs]{revtex4}

\usepackage{epsfig, amssymb}

\begin{document}
\advance\hoffset by  -4mm

\newcommand{\de}{\Delta E}
\newcommand{\mbc}{M_{\rm bc}}
\newcommand{\mds}{M(D_s)}
\newcommand{\bb}{B{\bar B}}
\newcommand{\qq}{q{\bar q}}
\newcommand{\kstar}{K^{*0}}
\newcommand{\kpi}{K^+\pi^-}
\newcommand{\kk}{K^+K^-}
\newcommand{\kkpi}{K^+K^-\pi^+}
\newcommand{\phipi}{\phi\pi^+}
\newcommand{\phik}{\phi K^-}
\newcommand{\kstark}{{\bar K^{*0}}K^+}
\newcommand{\ksk}{K_S^0 K^+}
\newcommand{\kspi}{K_S^0 \pi^+}
\newcommand{\kstarpip}{{\bar K^{*0}}\pi^+}
\newcommand{\kpipi}{K^-\pi^+\pi^+}
\newcommand{\bdsk}{{\bar B^0}\to D_s^+K^-}
\newcommand{\bdspi}{B^0\to D_s^+\pi^-}
\newcommand{\bdsstarpi}{B^0\to D^{*+}_s\pi^-}
\newcommand{\bdsh}{\bar{B^0}\to D_s^+ h^-}
\newcommand{\bdsstarh}{{\bar B^0}\to D^{*+}_s h^-}
\newcommand{\dpkpipi}{D^+\to\kpipi}
\newcommand{\dsphipi}{D_s^+\to\phipi}
\newcommand{\dskstark}{D_s^+\to\kstark}
\newcommand{\dsksk}{D_s^+\to\ksk}
\newcommand{\bdppi}{{\bar B^0}\to D^+\pi^-}
\newcommand{\br}{{\cal B}}

\title{\Large \rm 
Observation of $D_s^+K^-$ and evidence for $D_s^+\pi^-$ final states in
neutral $B$ decays}

\affiliation{Aomori University, Aomori}
\affiliation{Budker Institute of Nuclear Physics, Novosibirsk}
\affiliation{Chiba University, Chiba}
\affiliation{Chuo University, Tokyo}
\affiliation{University of Cincinnati, Cincinnati OH}
\affiliation{University of Frankfurt, Frankfurt}
\affiliation{Gyeongsang National University, Chinju}
\affiliation{University of Hawaii, Honolulu HI}
\affiliation{High Energy Accelerator Research Organization (KEK), Tsukuba}
\affiliation{Hiroshima Institute of Technology, Hiroshima}
\affiliation{Institute of High Energy Physics, Chinese Academy of Sciences, Beijing}
\affiliation{Institute of High Energy Physics, Vienna}
\affiliation{Institute for Theoretical and Experimental Physics, Moscow}
\affiliation{J. Stefan Institute, Ljubljana}
\affiliation{Korea University, Seoul}
\affiliation{Kyoto University, Kyoto}
\affiliation{Kyungpook National University, Taegu}
\affiliation{Institut de Physique des Hautes \'Energies, Universit\'e de Lausanne, Lausanne}
\affiliation{University of Ljubljana, Ljubljana}
\affiliation{University of Maribor, Maribor}
\affiliation{University of Melbourne, Victoria}
\affiliation{Nagoya University, Nagoya}
\affiliation{Nara Women's University, Nara}
\affiliation{National Kaohsiung Normal University, Kaohsiung}
\affiliation{National Lien-Ho Institute of Technology, Miao Li}
\affiliation{National Taiwan University, Taipei}
\affiliation{H. Niewodniczanski Institute of Nuclear Physics, Krakow}
\affiliation{Nihon Dental College, Niigata}
\affiliation{Niigata University, Niigata}
\affiliation{Osaka City University, Osaka}
\affiliation{Osaka University, Osaka}
\affiliation{Panjab University, Chandigarh}
\affiliation{Peking University, Beijing}
\affiliation{Princeton University, Princeton NJ}
\affiliation{RIKEN BNL Research Center, Brookhaven NY}
\affiliation{Saga University, Saga}
\affiliation{University of Science and Technology of China, Hefei}
\affiliation{Seoul National University, Seoul}
\affiliation{Sungkyunkwan University, Suwon}
\affiliation{University of Sydney, Sydney NSW}
\affiliation{Tata Institute of Fundamental Research, Bombay}
\affiliation{Toho University, Funabashi}
\affiliation{Tohoku Gakuin University, Tagajo}
\affiliation{Tohoku University, Sendai}
\affiliation{University of Tokyo, Tokyo}
\affiliation{Tokyo Institute of Technology, Tokyo}
\affiliation{Tokyo Metropolitan University, Tokyo}
\affiliation{Tokyo University of Agriculture and Technology, Tokyo}
\affiliation{Toyama National College of Maritime Technology, Toyama}
\affiliation{University of Tsukuba, Tsukuba}
\affiliation{Utkal University, Bhubaneswer}
\affiliation{Virginia Polytechnic Institute and State University, Blacksburg VA}
\affiliation{Yokkaichi University, Yokkaichi}
\affiliation{Yonsei University, Seoul}
  \author{P.~Krokovny}\affiliation{Budker Institute of Nuclear Physics, Novosibirsk} 
  \author{K.~Abe}\affiliation{High Energy Accelerator Research Organization (KEK), Tsukuba} 
  \author{K.~Abe}\affiliation{Tohoku Gakuin University, Tagajo} 
  \author{T.~Abe}\affiliation{Tohoku University, Sendai} 
  \author{I.~Adachi}\affiliation{High Energy Accelerator Research Organization (KEK), Tsukuba} 
  \author{Byoung~Sup~Ahn}\affiliation{Korea University, Seoul} 
  \author{H.~Aihara}\affiliation{University of Tokyo, Tokyo} 
  \author{M.~Akatsu}\affiliation{Nagoya University, Nagoya} 
  \author{Y.~Asano}\affiliation{University of Tsukuba, Tsukuba} 
  \author{T.~Aso}\affiliation{Toyama National College of Maritime Technology, Toyama} 
% ...
\collaboration{The Belle Collaboration}

\begin{abstract}
We report the first observation of a $B$ meson decay that is not
accessible by a direct spectator process. The channel $\bdsk$ is 
found in a sample of $85\times 10^6$ $\bb$ events,
collected with the Belle detector at KEKB, with a
branching fraction $\br(\bdsk)=(4.6^{+1.2}_{-1.1}\pm 1.3)\times 10^{-5}$.
We also obtain evidence for the $\bdspi$ decay with branching
fraction $\br(\bdspi)=(2.4^{+1.0}_{-0.8}\pm 0.7)\times 10^{-5}$.
This value may be used to extract a model-dependent value of $|V_{ub}|$.
\end{abstract}
\pacs{13.25.Hw, 14.40.Nd}
\maketitle

Although the $B$ mesons decay primarily through
the spectator processes, other processes such as
$W$-exchange and final state interactions (FSI)
may contribute appreciably,
especially~\cite{dsk_2} for rare modes that are now emerging at
the $B$ factory experiments.
As the focus of CP asymmetry studies shifts to rarer modes,
it is important to quantify the effects due to non-spectator and FSI
processes,
as they are often necessary for generating measurable CP asymmetries.
Since the non-spectator effects carry significant theoretical
uncertainties, a quantitative understanding must be developed
before we can extract CKM matrix parameters.

Important insight in this regard can be gained by experimental measurement
of channels where non-spectator processes dominate.
We report here the first observation of a mode that is not directly
accessible through the spectator process.
The mode $\bdsk$ may occur via $W$-exchange or final state
rescattering, and predictions for its branching fraction vary
over a wide range, $3\times
10^{-6}-10^{-4}$~\cite{dsk_2,dsk_1,dsk_3,dsk_4}.
The search for $\bdsk$ also encompasses $\bdspi$, a mode that is
expected to be dominated by a (spectator) $b\to u$ transition.
As it lacks a penguin contribution, it can in principle
provide a way to determine $|V_{ub}|$~\cite{vub2}.

The results reported here are based on a 78.7~fb$^{-1}$
data sample, collected with the Belle detector~\cite{NIM} at the KEKB
asymmetric energy
$e^+e^-$ collider~\cite{KEKB} at the center-of-mass (CM) energy of the
$\Upsilon(4S)$ resonance and
containing $85.0\times 10^6$ produced
$B\bar B$ pairs.
A 7.5~fb$^{-1}$ data sample taken at a CM energy that
is 60 MeV below the $\Upsilon(4S)$ resonance is used
for systematic studies of the $e^+e^-\to q\bar q$ background.


The Belle detector has been described elsewhere\cite{NIM}.
Charged tracks are selected with requirements based on the
average hit residual and impact parameter relative to the
interaction point (IP). We also require that the transverse momentum of
the tracks be greater than 0.1 GeV$/c$ in order to reduce the low 
momentum combinatorial background.
For charged particle identification (PID) the combined information
from specific ionization in the central drift chamber ($dE/dx$), time-of-flight scintillation counters (TOF) and aerogel \v{C}erenkov counters (ACC) is used.
At large momenta ($>2.5$~GeV$/c$) only the ACC and $dE/dx$ are used.
Charged kaons are selected with PID criteria that have
an efficiency of 88\%, a pion misidentification probability of 8\%,
and negligible contamination from protons.
The criteria for charged pions have an efficiency of 89\% and 
a kaon misidentification probability of 9\%.
All tracks that are positively identified as electrons are rejected.

Neutral kaons are reconstructed via the decay $K_S^0\to\pi^+\pi^-$.
The two-pion invariant mass is required to be within 6~MeV$/c^2$ 
($\sim 2.5\sigma$) of the nominal $K^0$ mass, and the displacement of 
the $\pi^+\pi^-$ vertex from the IP in the transverse 
$r$-$\phi$ plane is required to be between 0.1~cm and 20~cm. 
The directions in the $r$-$\phi$ projection of the $K_S^0$ candidate's flight path and momentum are required to agree within 0.2 radians.

We reconstruct $D_s^+$ mesons in the channels
$D_s^+\to \phipi$, $\kstark$, and $\ksk$ (inclusion of charge conjugate states is implicit throughout this report).
$\phi$ ($\kstar$) mesons are formed from the $\kk$ ($\kpi$ ) pairs with
invariant mass within 10~MeV$/c^2$ (50~MeV$/c^2$) of the nominal $\phi$ ($\kstar$) mass.
We select $D_s^+$ mesons in a wide ($\pm 0.5$~GeV$/c^2$) window,
for subsequent studies; the $\mds$
signal region is defined to be within 12~MeV$/c^2$ ($\sim 2.5\sigma$)
of the nominal $D_s^+$ mass. $D_s^+$ candidates are combined with a 
charged kaon or pion to form a $B$ meson.
Candidate events are identified by their CM
energy difference, \mbox{$\de=(\sum_iE_i)-E_b$}, and the
beam constrained mass, $\mbc=\sqrt{E_b^2-(\sum_i\vec{p}_i)^2}$, where
$E_b = \sqrt{s}/2$ is the beam energy and 
$\vec{p}_i$ and $E_i$ are the momenta and
energies of the decay products of the $B$ meson in the CM frame.
We select events with $\mbc>5.2$~GeV$/c^2$
and $|\de|<0.2$~GeV and define the $B$ signal region to be
$5.272$~GeV$/c^2<\mbc<5.288$~GeV$/c^2$ and $|\de|<0.03$~GeV.
The $\mbc$ sideband is defined as $5.20$~GeV$/c^2<\mbc<5.26$~GeV$/c^2$.
We use a Monte Carlo (MC) simulation to determine the efficiency~\cite{GEANT}.


To suppress the large combinatorial background that is dominated by 
the two-jet-like $e^+e^-\to\qq$ 
continuum 
process, variables that characterize the event topology are used. 
We require $|\cos\theta_{\rm thr}|<0.80$, where $\theta_{\rm thr}$ is 
the angle between the thrust axis of the $B$ candidate and that of the 
rest of the event.  This requirement eliminates 77\% of the continuum 
background and retains 78\% of the signal events. We also define a 
Fisher discriminant, ${\cal F}$, 
that includes
the production angle of 
the $B$ candidate, the angle of the $B$ candidate thrust axis with 
respect to the beam axis, and nine parameters that characterize the 
momentum flow in the event relative to the $B$ candidate thrust axis 
in the CM frame~\cite{VCal}. We impose a requirement on ${\cal{F}}$ 
that rejects 50\% of the remaining continuum background and 
retains 92\% of the signal.

We also consider possible backgrounds from $\qq$ events containing
real $D_s^+$ mesons.  These events peak in the $\mds$ spectra
but not in the $\de$ and $\mbc$ distributions. 
We study this background using the $\mbc$ sideband and find
it to be fewer than 0.1 and 0.5 events for the $\bdsk$ and $\bdspi$ 
modes, respectively. 

\begin{figure}
  \includegraphics[width=0.5\textwidth] {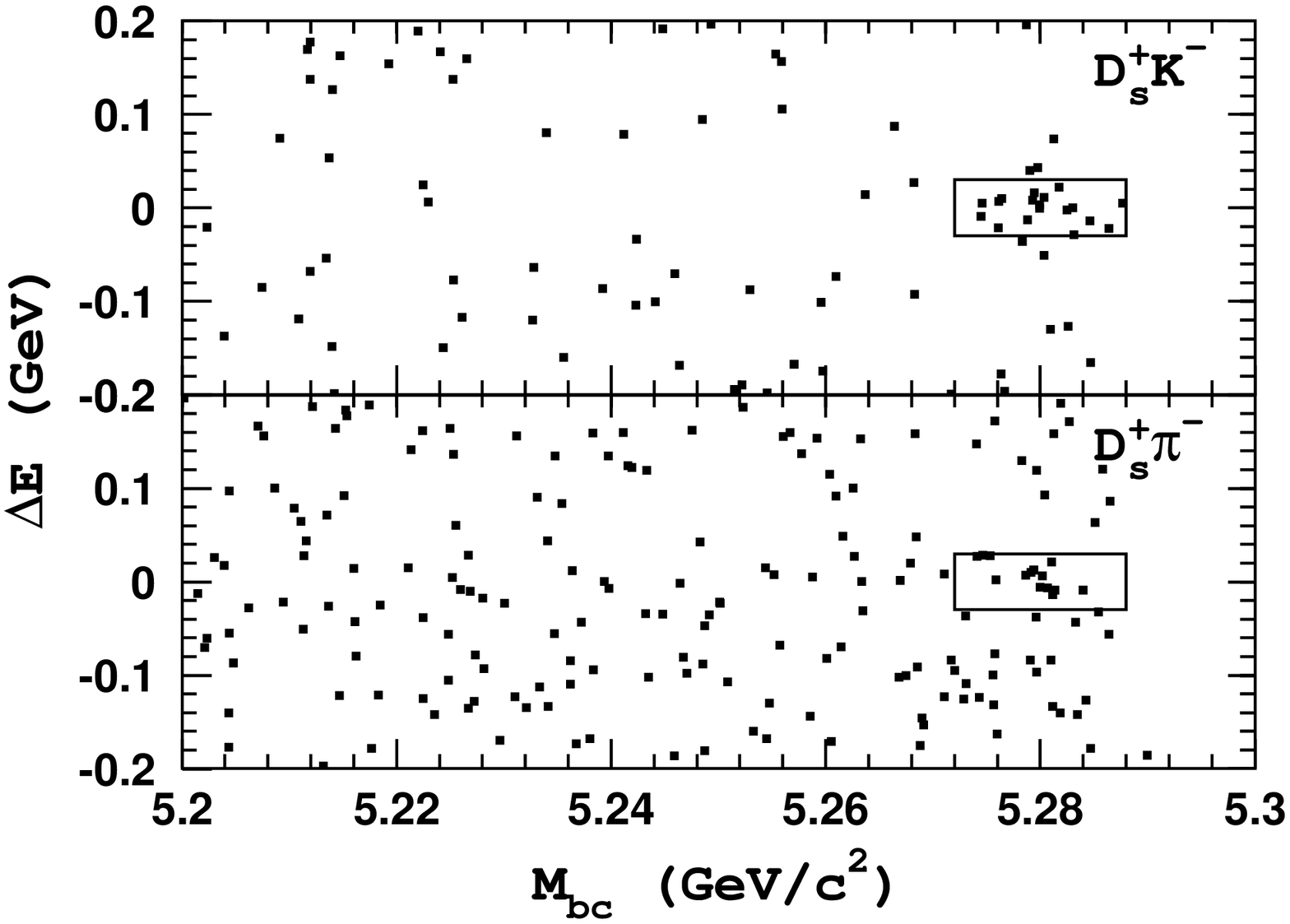} 
  \caption{The $\de$ versus $\mbc$ scatter plot for the $\bdsk$ (top) and
    $\bdspi$ (bottom) candidates in the $\mds$ signal region.
  The points represent the
  experimental data and the boxes show the $B$ meson signal region.}
  \label{dsh_mbcde}
\end{figure}

\begin{figure}
  \includegraphics[width=0.5\textwidth] {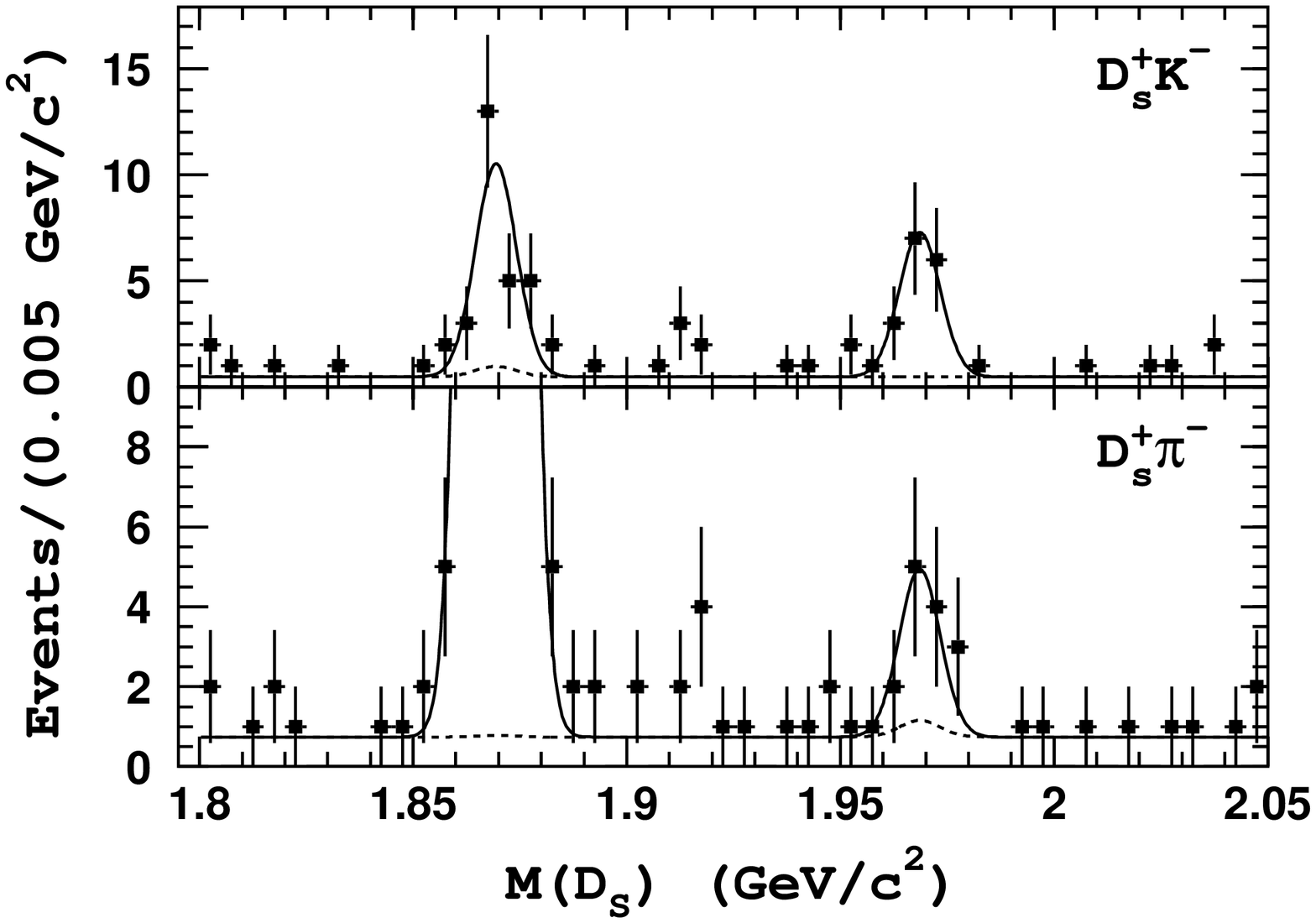} 
  \caption{The $\mds$ spectra for $\bdsk$ (top) and
    $\bdspi$ (bottom) in the $B$ signal region. 
    The points with errors represent 
    experimental data and the curves display the results of 
    the simultaneous fit described in the text.
}
  \label{dsh_mds}
\end{figure}

Other $B$ decays, such as $\bdppi$, $\dpkpipi$, with one pion 
misidentified as a kaon, require particular attention because they 
have large branching fractions and can peak in 
the $\mbc$ signal region. 
The reconstructed invariant mass spectra for these events
overlap with the signal $D_s^+$ mass region, while their $\de$ 
distribution is shifted by about 50~MeV$/c^2$.
To suppress this background, we exclude event candidates 
that are consistent with the $\dpkpipi$ mass hypothesis within 
15~MeV$/c^2$ ($\sim 3\sigma$) when the two same-sign particles
are considered to be pions, independently of their PID information.
For the $\dsksk$ mode there is a similar background from $\bdppi$, $D^+\to\kspi$. In this case we
exclude candidates consistent within 20~MeV$/c^2$ ($\sim 3\sigma$)
with the $D^+\to\kspi$ hypothesis. 

Possible backgrounds from  $B$ decays via $b\to c$ transitions 
($B\to D_s D X$) are also considered.
The $D_s^+$ from these decays have a lower momentum and 
are kinematically separated from the signal.
We analyzed a MC sample of generic $\bb$ events corresponding to about twice the data sample and found
no peaking backgrounds.

Another potential $\bb$ background is charmless 
${\bar B^0}\to K^-K^+K^-\pi^+ (K_S^0\kk)$. Such events peak in the $\de$ and $\mbc$ spectra, but not in the $\mds$ distributions. 
They tend to be dominated by quasi-two-body decay 
channels such as $\phi {\bar K^{(*)0}}$~\cite{phikstar, kshh}.
To reduce this background, we reject events with low 
($<2$~GeV$/c^2$) two particle invariant masses: $M_{K^-\pi^+}$ and 
$M_{\phik}$ for the $\dsphipi$ channel, 
$M_{\kk}$ and $M_{{\bar K^{*0}}K^-}$ for $\dskstark$, 
and $M_{\kk}$ and $M_{K_S^0 K^-}$ for $\dsksk$.
The remaining background from these sources, if any, is excluded by 
fitting the $\mds$ distribution.



\begin{table*}
\caption{Results on the signal yields and branching fractions.
The efficiencies do not include intermediate branching fractions.}
\footnotesize
\medskip
\label{defit}
  \begin{tabular*}{\textwidth}{l@{\extracolsep{\fill}}cccccc}\hline\hline
 Mode  & $\mds$ - $\de$ yield & $\mds$ yield & $\de$ yield 
& Efficiency, \% & 
${\cal B}$ $(10^{-5})$ & 
Stat. significance\\\hline
$\bdsk$, $\dsphipi$ & $8.9^{+3.3}_{-2.7}$ &
        $8.9^{+3.4}_{-2.7}$ & $9.0^{+3.3}_{-2.7}$ &
        $11.6\pm 0.4$ & $5.1^{+2.0}_{-1.6}\pm 1.4$ & $5.1\sigma$\\

$\bdsk$, $\dskstark$ & $6.1^{+3.0}_{-2.3}$ &
        $5.1^{+2.8}_{-2.2}$ & $5.9^{+3.0}_{-2.4}$ &
        $6.8\pm 0.3$ & $4.8^{+2.4}_{-1.8}\pm 1.3$ & $3.8\sigma$\\

$\bdsk$, $\dsksk$ &  $1.6^{+2.0}_{-1.2}$ &
        $1.0^{+1.9}_{-1.0}$ & $2.8^{+2.3}_{-1.6}$ &
        $7.0\pm 0.3$ & $2.2^{+2.8}_{-1.7}\pm 0.6$ & $1.6\sigma$\\
\hline
$\bdsk$, simultaneous fit & $16.4^{+4.6}_{-3.9}$ &
        $15.0^{+4.5}_{-3.8}$ & $17.5^{+4.8}_{-4.2}$ & --- &
        $4.6^{+1.2}_{-1.1}\pm 1.3$ & $6.4\sigma$\\\hline
$\bdspi$, $\dsphipi$ & $4.7^{+2.6}_{-2.0}$ &
        $4.8^{+2.6}_{-1.9}$ & $4.0^{+2.6}_{-2.0}$ & 
        $12.9\pm 0.4$ & $2.4^{+1.3}_{-1.0}\pm 0.7$ & $3.2\sigma$\\

$\bdspi$, $\dskstark$ & $3.4^{+3.2}_{-2.4}$ &
        $2.9^{+2.8}_{-2.0}$ & $4.4^{+3.3}_{-2.6}$ & 
        $7.5\pm 0.3$ & $2.4^{+2.3}_{-1.7}\pm 0.7$ & $1.6\sigma$\\

$\bdspi$, $\dsksk$ & $1.6^{+2.3}_{-1.6}$ &
        $2.2^{+2.2}_{-1.5}$ & $0.9^{+2.2}_{-0.9}$ &
        $7.2\pm 0.3$ & $2.2^{+3.1}_{-2.2}\pm 0.6$ & $0.9\sigma$\\
\hline
$\bdspi$, simultaneous fit & $10.1^{+4.4}_{-3.7}$ &
        $10.3^{+4.1}_{-3.4}$ & $9.5^{+4.5}_{-3.8}$ & --- &
        $2.4^{+1.0}_{-0.8}\pm 0.7$ & $3.6\sigma$\\\hline\hline
  \end{tabular*}
\end{table*}

The scatter plots in  $\de$ and $\mbc$  for the $\bdsk$ and $\bdspi$ 
candidates in the $\mds$ signal region are shown in
Fig.~\ref{dsh_mbcde}; a significant enhancement in the 
$B$ signal region is observed.
Figure~\ref{dsh_mds} shows the $\mds$ spectra for selected 
$\bdsk$ and $\bdspi$ candidates in the $B$ signal region.
In addition to clear signals at the $D_s^+$ mass in
Fig.~\ref{dsh_mds}, we  observe peaks at the $D^+$ mass, 
corresponding to $\bdppi$ and ${\bar B^0}\to D^+ K^-$, 
$D^+\to\phipi, \kstark, \ksk$. 

Our studies have shown that the backgrounds may peak in the signal
region of $\mds$ or of $\de$ (and $\mbc$) but not in both
simultaneously. To extract our signal, we therefore perform a binned
maximum likelihood fit to the two-dimensional distribution of data in
$\mds$ and $\de$, separating the backgrounds from the signal
component, which peaks in both.
For each of the three $D_s^+$ decay channels the $\de$ range,
$-0.1$~GeV$<\de<0.2$~GeV, is divided into 30 bins and the $\mds$
range, $1.5<\mds<2.5$~GeV$/c^2$, into 200  bins.
All bins in all $D_s^+$ submodes are fitted simultaneously
to a sum of signal and background shapes.
The $D_s^+$ signal is described by a two-dimensional Gaussian, with
widths in both dimensions obtained and fixed using reconstructed
signals in the data from $\bdppi (D^+\to K^-\pi^+\pi^+, K_S \pi^+)$.
The signal amplitude is constrained to correspond to the same
branching fraction $\br(\bdsh)$ for all three $D_s^+$ submodes.
The fit also includes an additional two-dimensional Gaussian 
for ${\bar B^0}\to D^+h^-$ decays.

\begin{figure}
  \includegraphics[width=0.5\textwidth] {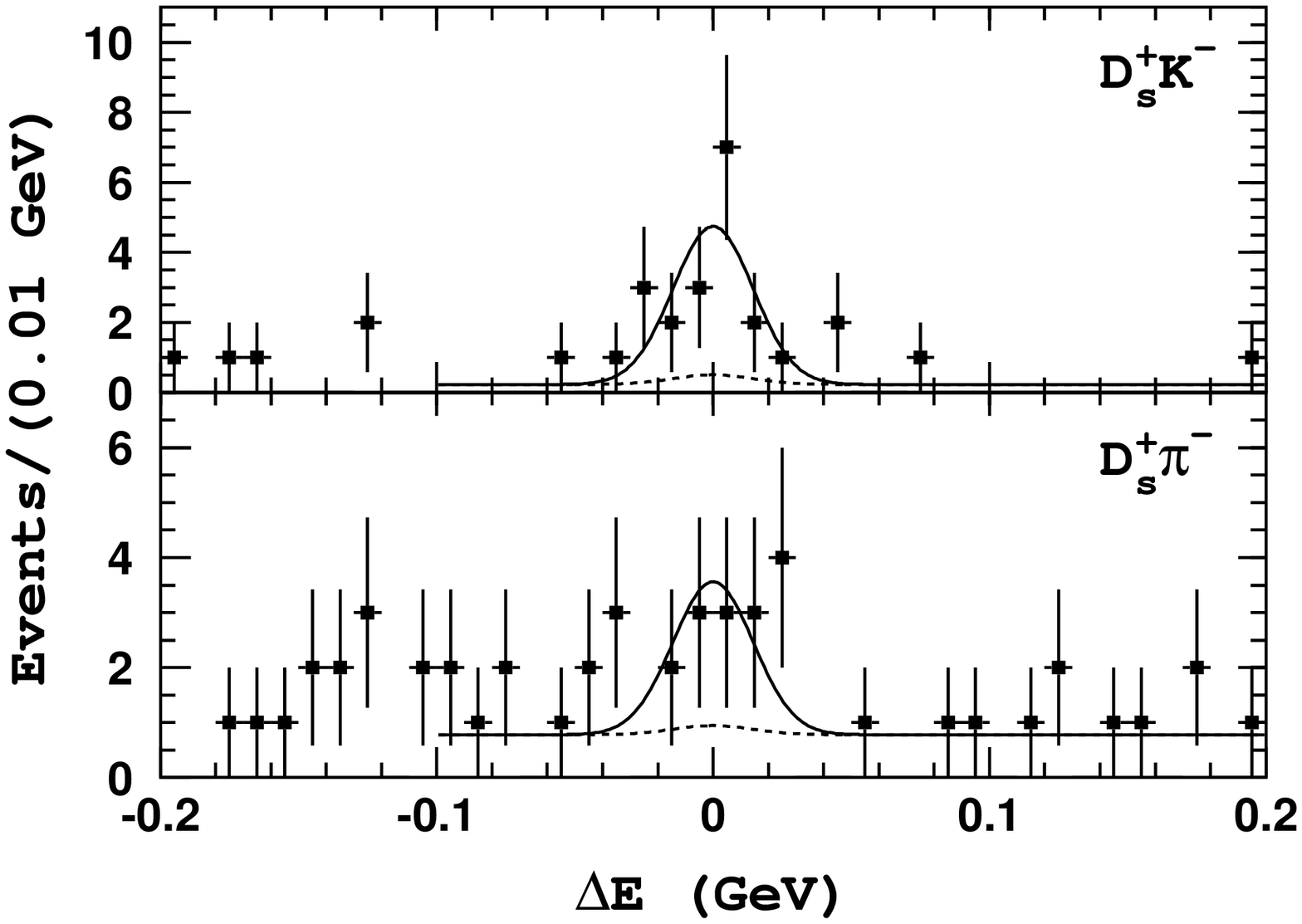}
  \caption{The $\de$ spectra for  
$\bdsk$ (top) and $\bdspi$ (bottom)  in the $B$ signal region. The points with errors are 
    experimental data and the curves are the results of 
    the simultaneous fit described in the text.
}
  \label{dsh_mdsde}
\end{figure}

The background includes three components: combinatorial (flat in 
$\mds$ and $\de$), $\qq$ events that peak in $\mds$ and are flat in 
$\de$, and $B$ decays that peak in $\de$ and are flat in $\mds$.
The levels of the three  components are allowed to vary 
independently in the three reconstructed $D_s^+$ modes.
The fit results are given in Table~\ref{defit}.
The statistical significance quoted in Table~\ref{defit} is defined as 
$\sqrt{-2\ln ({\cal L}_0/{\cal L}_{max})}$, where ${\cal L}_{max}$ and 
${\cal L}_{0}$ denote the maximum likelihood with the fitted signal 
yield and with the signal yield fixed to zero, respectively.
The results of one-dimensional fits to the $\mds$ and $\de$ 
distributions are also shown in Table~\ref{defit} for comparison.
Figures~\ref{dsh_mds} and \ref{dsh_mdsde} show the $\mds$ and $\de$
projections for events from the signal region and the fitted 
signal plus background combined shape by solid lines and
background shape including the peaking background by dashed lines.
The peaking background is found to be 
$1.0\pm 0.5$ and $1.6\pm 1.0$ events for $\bdsk$ and $\bdspi$, respectively. 

$\bdsstarh$ final states,  where the low energy photon from the 
$D^*_s\to D_s\gamma$ decay is missed,
can populate the $\bdsh$ signal region. 
These  would produce a long tail on the negative side of the $\de$ 
distribution.  
In theoretical models based on factorization, 
the $\bdsstarpi$ and $\bdspi$ decay widths are predicted to be 
approximately equal; there are, however, no 
corresponding predictions  for ${\bar B^0}\to D^{(*)+}_s K^-$ decays.
To study the sensitivity of the measured branching fraction to a 
possible $\bdsstarh$ contribution, we perform a fit with an additional 
$\bdsstarh$ component included, where the signal shape is fixed from the 
MC and the branching fraction is left as a free parameter. The resulting
2\% difference in the $\bdspi$ event yield (compared to the results 
presented in Table~\ref{defit}) is added to the systematic
uncertainty; the change in the $\bdsk$ yield is less than 1\%.
We also check for crossfeed between $\bdsk$ and $\bdspi$
due to kaon/pion misidentification.
To study this we include the crossfeed contributions in the
simultaneous fit, with shapes 
fixed from the MC and misidentification
probabilities obtained from data; 
the uncertainty due to this effect
is found to be  negligible ($\lesssim 1\%$).

As a check, we apply the same procedure to 
$\bdppi$ and ${\bar B^0}\to D^+ K^-$, $D^+\to\phipi, \kstark, \ksk$, 
and obtain 
$\br(\bdppi)=(2.8\pm 0.2)\times 10^{-3}$ and 
$\br({\bar B^0}\to D^+ K^-)=(3.0\pm 0.7)\times 10^{-4}$,
which agree well with the PDG values 
$\br(\bdppi)=(3.0\pm 0.4)\times 10^{-3}$ and
$\br({\bar B^0}\to D^+ K^-)=(2.0\pm 0.6)\times 10^{-4}$~\cite{PDG}.

The following sources of systematic error are found to be 
the most significant: tracking efficiency (2\% per track), charged hadron identification 
efficiency (2\% per particle), $K^0_S$ reconstruction efficiency (6\%),
signal-shape parameterization (5\%) and MC statistics (3\%).
The tracking efficiency error is estimated using
$\eta$ decays to $\gamma\gamma$ and $\pi^+\pi^-\pi^0$.
The $K/\pi$ identification uncertainty is determined 
from $D^{*+}\to D^0\pi^+$, $D^0\to K^-\pi^+$ decays.
We assume equal production of $B^+B^-$ and $B^0\bar B^0$ pairs but do
not include an additional error from this
assumption.
The uncertainty in the $D_s^+$ meson branching fractions, which is 
dominated by the 25\% error on $\br(\dsphipi)$, is also taken
into account.
The overall systematic uncertainty is 28\%.


In summary, we report the first observation of  $\bdsk$
with a $6.4\sigma$ statistical significance. We find
$\br(\bdsk)=(4.6^{+1.2}_{-1.1}\pm 1.3)\times 10^{-5}$, which is
consistent with a calculation of the $W$-exchange rate in
the ``PQCD factorization" approach~\cite{dsk_3},
but much higher than an earlier result~\cite{dsk_1}.
On the other hand, it should be noted that
the recent observation of higher-than-predicted rates
for $\bar{B}^0\to D^0h^0$ ($h^0 = \pi^0,\ \eta,\
\omega$)~\cite{dpi0_belle,dpi0_cleo,dpi0_babar}
and $\bar{B}^0\to D^0\rho^0$~\cite{drho_belle} 
suggest that FSI may contribute appreciably to $\bdsk$~\cite{dsk_2,dsk_4}.
We also obtain evidence for $\bdspi$ with
$\br(\bdspi)=(2.4^{+1.0}_{-0.8}\pm 0.7)\times 10^{-5}$
($3.6\sigma$ statistical significance).
Our results are consistent with recent evidence
from BaBar~\cite{dspi_babar}.

Since the dominant systematic uncertainty on both  measurements is due to
the branching fraction of $\dsphipi$, $\br_{\phi\pi}$, we also report
$\br(\bdsk)\times\br_{\phi\pi}=
(16.4^{+4.5}_{-3.8}\pm 2.1)\times 10^{-7}$ and
$\br(\bdspi)\times\br_{\phi\pi}=
(8.6^{+3.7}_{-3.0}\pm 1.1)\times 10^{-7}$.
Using $\br(\bdspi)/\br(B^0\to D_s^+ D^-) = (0.424\pm 0.041)\times
|V_{ub}/V_{cb}|^2$~\cite{vub2},
and $\br(B^0\to D_s^+ D^-)\times\br_{\phi\pi}=(3.0\pm 1.1)\times 10^{-4}$
calculated from a CLEO result~\cite{dsd_cleo},
we can extract a model-dependent value
$|V_{ub}/V_{cb}| = (8.2^{+3.5}_{-2.9}\pm 3.4)\times 10^{-2}$,
where no error on the factorization assumption or other sources of
model-dependence is included.
This  value is in agreement with ones obtained from inclusive
semileptonic decays\cite{PDG}.

We wish to thank the KEKB accelerator group for the excellent
operation of the KEKB accelerator.
We acknowledge support from the Ministry of Education,
Culture, Sports, Science, and Technology of Japan
and the Japan Society for the Promotion of Science;
the Australian Research Council
and the Australian Department of Industry, Science and Resources;
the National Science Foundation of China under contract No.~10175071;
the Department of Science and Technology of India;
the BK21 program of the Ministry of Education of Korea
and the CHEP SRC program of the Korea Science and Engineering Foundation;
the Polish State Committee for Scientific Research
under contract No.~2P03B 17017;
the Ministry of Science and Technology of the Russian Federation;
the Ministry of Education, Science and Sport of the Republic of Slovenia;
the National Science Council and the Ministry of Education of Taiwan;
and the U.S. Department of Energy.

\end{document}